\documentclass[a4paper,fleqn,usenatbib]{mnras}

\usepackage[T1]{fontenc}
\usepackage{ae,aecompl}

\usepackage{graphicx}	% Including figure files
\usepackage{amsmath}	% Advanced maths commands
\usepackage{amssymb}	% Extra maths symbols

\def\gtsima{$\; \buildrel > \over \sim \;$}
\def\simgt{\lower.5ex\hbox{\gtsima}}

\def\songmei2022{https://doi.org/10.48550/arXiv.2212.11034}

\title[Open cluster chemical tagging]{Recovering extra-tidal open cluster members via multi-elemental chemical tagging}

\author[Andr\'es E. Piatti]{
Andr\'es E. Piatti$^{1,2}$\thanks{E-mail: andres.piatti@fcen.uncu.edu.ar} \\
$^{1}$Instituto Interdisciplinario de Ciencias B\'asicas (ICB), CONICET-UNCuyo, 
Padre J. Contreras 1300, M5502JMA, Mendoza, Argentina\\
$^{2}$Consejo Nacional de Investigaciones Cient\'{\i}ficas y T\'ecnicas, Godoy Cruz 
2290, C1425FQB,  Buenos Aires, Argentina\\}

\date{Accepted XXX. Received YYY; in original form ZZZ}

\pubyear{}

\begin{document}
\label{firstpage}
\pagerange{\pageref{firstpage}--\pageref{lastpage}}
\maketitle

\begin{abstract}
The identification of open cluster (OC) members has been revolutionized by high-precision 
{\it Gaia} astrometry, yet traditional kinematic membership selections remain inherently 
conservative, often overlooking stars in tidal tails or those with perturbed velocities. 
This study investigates the reliability of these kinematic probabilities by searching for 
leaky cluster members -- stars that fail standard kinematic membership criteria 
($P < 0.7$) but possess chemical signatures identical to their host clusters. Using 
high-resolution spectroscopic data from the {\it Gaia}-ESO Survey, we established a 
seven-element chemical fingerprint ([Fe/H], Li, Si, Ca, Ti, Co, and Ni) for 34 OCs. We 
identified a sample of 63 stars across 22 clusters that are chemically indistinguishable 
from their host populations despite being kinematically rejected by standard algorithms. 
By cross-referencing these targets with Jacobi radii ($r_J$) derived from modern Milky 
Way potential models, we find that 35\% are located in extra-tidal regions ($d > r_J$), 
providing direct evidence of active cluster dissolution and tidal debris. The remaining 
65\% are located within the Jacobi radius, suggesting that their kinematic rejection is 
likely due to orbital motion in unresolved binary systems. These results demonstrate that 
chemical tagging is a critical tool for overcoming the spatial and kinematic biases of 
astrometric catalogues. By recovering these lost members, we provide a more complete census of 
cluster mass loss and underscore the necessity of a hybrid chemical-kinematic 
approach to map the transition of stars from bound systems to the Galactic field.

\end{abstract}


%%%%%%%%%%%%%%%%%%%%%%%%%%%%%%%%%%%%%%%%%%%

%%%%%%%%%%%%%%%%% BODY OF PAPER %%%%%%%%%%%%%%%%%

\section{Introduction}  

The dissolution of open clusters (OCs) represents one of the primary mechanisms 
for the assembly and enrichment of the Milky Way thin disk. Formed 
within giant molecular clouds, these systems represent a snapshot of the chemical and 
dynamical conditions of their birth environments. However, OCs are transient structures; 
throughout their lifetimes, they are subject to continuous mass loss due to internal 
two-body relaxation, stellar evolution, and external perturbations from the Galactic tidal 
field and encounters with giant molecular clouds \citep{spitzer1987,bm2003}. Identifying these escaped stars is crucial for reconstructing 
the star formation history of the Milky Way and understanding the origin of the field
star population.

The advent of the {\it Gaia} mission has transformed the field of cluster dynamics, providing 
high-precision astrometry and membership probabilities ($P$) for thousands of systems
\citep{cantatgaudinetal2020,hr2024}. Indeed, our ability to identify cluster members has 
improved exponentially. High-precision astrometry allows for the determination of membership
probabilities based on 5D or 6D phase-space distributions (e.g., UPMASK, HDBSCAN). 
However, these kinematic assignments are fundamentally conservative. Standard membership 
algorithms, such as those employed by \citet{cantatgaudinetal2020} and \citet{hr2024},
typically define high-confidence members as those with $P$ $>$ 0.7 or $P$ $>$ 0.9. 
While this maintains sample purity, it introduces a significant selection bias. 
Stars in the high-velocity tails of the Maxwellian distribution, unresolved binary systems
-- where orbital motion creates a velocity offset from the cluster mean -- and stars already 
residing in tidal tails are frequently assigned low membership probabilities. 
These kinematic outliers are often discarded as field interlopers, even when they remain physically associated 
with the cluster's evolutionary history.

Dynamically, the boundary of a cluster is best defined by its Jacobi radius, the point at 
which the cluster’s gravitational pull is balanced by the Galactic tidal field. 
Stars crossing this boundary into the extra-tidal regime begin to form tidal tails, trailing and leading the 
cluster along its orbit. Recent works, such as those by \citet{hr2024}, have provided a 
rigorous framework for determining these radii by accounting for the cluster mass and the local 
Galactic potential. Identifying stars that have crossed this threshold but retain their cluster 
identity is a piece of evidence for observing cluster dissolution in real-time.

Chemical tagging -- the hypothesis that stars born from the same molecular cloud share a 
unique chemical footprint \citep{fbh2002} -- offers a powerful independent 
diagnostic to recover these lost members. Unlike kinematic signatures, which can be altered by dynamical encounters or tidal forces, the atmospheric abundances of slow-evolving stars remain a permanent 
record of their birth environment \citep{desilvaetal2007,mitschangetal2014,blancoetal2015,bovy2016,
donoretal2020}. While recent studies have suggested that the homogeneity of the interstellar medium over 
the last billion years may lead to significant chemical overlap between different open clusters 
\citep[e.g.][]{casamiquelaetal2021}, the use of high-dimensional chemical spaces and targeted, 
supervised tagging remains a robust method for distinguishing cluster members from the field 
population. Recent high-resolution spectroscopic surveys, such as the 
{\it Gaia}-ESO Survey \citep{gilmoreetal2012} and APOGEE \citep{majewskietal2017}, 
have begun to provide the multi-elemental precision necessary to distinguish true cluster 
members from field interlopers with similar kinematics. Indeed, by leveraging this 
high-resolution spectroscopy, we can identify stars that are chemically indistinguishable from 
a cluster’s core population, regardless of their current kinematic state or spatial position.

In this work, we present a systematic analysis of stars that are kinematically rejected by 
standard algorithms ($P < 0.7$) but are chemically consistent with their parent clusters. 
We utilize a seven-element chemical fingerprint ([Fe/H], Li, Si, Ca, Ti, Co, and Ni) to investigate 34 open clusters. By
mapping these stars relative to the Jacobi radii provided by \citet{hr2024}, we
demonstrate that chemical tagging successfully recovers two critical populations: (1) 
extra-tidal escapees residing in tidal debris, and (2) internal outliers, such as binary systems, 
whose kinematic probabilities have been underestimated. This approach allows for a more complete
census of cluster populations and provides a clearer view of the leaky nature of open clusters
in the Milky Way disk. We quantify the kinematic bias inherent in modern catalogues and 
demonstrate that chemical tagging is essential for mapping the full spatial and dynamical 
extent of the Milky Way's cluster population. 
In Section~2 we describe the employed data and the selection of open cluster members. 
We analyze and discuss our findings in Section~3, and in Section~4 we summarize the main 
conclusions of this work.

\section{Data handling}

The primary data for this study were drawn from the {\it Gaia}-ESO Survey 
\citep[GES;][]{gilmoreetal2012,randichetal2013}. GES is a high-resolution, public 
spectroscopic survey conducted with the FLAMES multi-object spectrograph on the 
Very Large Telescope (VLT) at Paranal Observatory. The survey was designed to provide 
a homogeneous overview of the kinematics and elemental abundances of all major components 
of the Milky Way, with a significant portion of its time dedicated to the study of open 
clusters across all ages and Galactic locations. 

The GES data provide a unique advantage for this study due to its use of the GIRAFFE and 
UVES spectrographs, which allow for the measurement of precise radial velocities and 
detailed chemical abundances for a wide variety of elements, ranging from light elements 
(e.g., Li) to iron-peak and $\alpha$-elements for 114324 selected stars.

To ensure the reliability of our chemical tagging analysis and the robustness of the identified
outliers, we applied a series of stringent quality cuts to the initial GES catalogue. These 
criteria were designed to minimize the impact of observational noise and physical parameters 
that could lead to degenerate or inaccurate abundance determinations. Following the 
methodology established in \citet{boucheretal2026}, the selection criteria are as follows: 

-- Signal-to-Noise ratio (S/N): we restricted the sample to stars with S/N $>$ 20. This 
threshold is necessary to ensure that the spectral lines for our target elements 
(Fe,Li,Si,Ca,Ti,Co,Ni) are sufficiently well-defined for precise abundance extraction, 
particularly for the weaker lines of iron-peak elements.

-- Effective Temperature ($T_{eff}$): we selected stars within the temperature range of 
3000 K to 7000 K. This range focuses on the most reliable part of the spectroscopic 
pipelines, excluding very hot stars where non-LTE effects and line broadening become dominant, 
as well as extremely cool stars where molecular blending can severely complicate the continuum 
placement.

-- Target Classification: we  only utilized  the \texttt{GES\_TYPE} flag to specifically 
target stars associated with open cluster fields. Only stars labeled as \texttt{*\_OC} (open cluster) 
or \texttt{*\_CL} (cluster) were included. This ensures that our starting sample consists of stars 
that are either confirmed members or candidates located within the spatial footprint of 
known stellar systems. 

After applying these filters, we obtained a high-quality final sample consisting of 6050 stars
in 60 open clusters. This dataset includes open cluster and star names, RA and Dec coordinates,
radial velocities, membership probability, stellar atmospheric parameters, and abundances for 
the 31 different chemical species included in GES.

\begin{table}
\tiny
\caption{Number of stars measured in each selected open cluster and chemical element.}
\label{tab1}
\begin{tabular}{lccccccc}
\hline
Name & [Fe/H] & Li & Si & Ca &  Ti & Co & Ni \\
     & (dex) & (dex) & (dex) & (dex) & (dex) & (dex) & (dex) \\
\hline
Blanco~1 & 41 & 40 & 16 & 16 & 16 & 16 & 16\\
Berkeley~21 & 81 & 67 & 22 & 43 & 27 & 27 & 36\\
Berkeley~31 & 77 & 74 & 73 & 74 & 73 & 73 & 73\\
Berkeley~32 & 174 & 142 & 58 & 150 & 59 & 32 & 116\\
Berkeley~36 & 114 & 114 & 107 & 103 & 107 & 106 & 107\\
Berkeley~39 & 351 & 345 & 48 & 98 & 35 & 26 & 74\\
Berkeley~44 & 42 & 35 & 16 & 14 & 16 & 15 & 17\\
Berkeley~81 & 39 & 29 & 18 & 18 & 18 & 18 & 18\\
Haffner~10 & 191 & 183 & 25 & 112 & 24 & 25 & 47\\
IC~2602 & 18 & 17 & 11 & 11 & 11 & 11 & 11\\
NGC~2141 & 395 & 391 & 57 & 288 & 150 & 52 & 86\\
NGC~2158 & 243 & 223 & 60 & 109 & 94 & 59 & 73\\
NGC~2243 & 338 & 331 & 325 & 332 & 324 & 325 & 330\\
NGC~2264 & 50 & 43 & 14 & 14 & 14 & 15 & 12\\
NGC~2420 & 315 & 312 & 284 & 313 & 285 & 265 & 290\\
NGC~2425 & 124 & 124 & 21 & 58 & 26 & 16 & 26\\
NGC~2516 & 356 & 351 & 181 & 151 & 174 & 167 & 198\\
NGC~2547 & 34 & 33 & 17 & 16 & 17 & 17 & 23\\
NGC~2682 & 115 & 110 & 114 & 114 & 114 & 88 & 114\\
NGC~3532 & 429 & 393 & 221 & 171 & 103 & 105 & 291\\
NGC~4815 & 35 & 33 & 14 & 16 & 13 & 14 & 14\\
NGC~6005 & 67 & 35 & 28 & 41 & 36 & 35 & 32\\
NGC~6067 & 76 & 68 & 27 & 57 & 35 & 23 & 43\\
NGC~6281 & 44 & 27 & 17 & 32 & 29 & 23 & 21\\
NGC~6253 & 279 & 145 & 198 & 212 & 212 & 208 & 197\\
NGC~6259 & 41 & 40 & 23 & 32 & 22 & 22 & 24\\
NGC~6405 & 77 & 60 & 36 & 62 & 54 & 18 & 53\\
NGC~6633 & 27 & 19 & 13 & 20 & 19 & 12 & 15\\
NGC~6705 & 181 & 147 & 63 & 113 & 91 & 62 & 75\\
NGC~6802 & 47 & 37 & 17 & 20 & 18 & 16 & 17\\
Ruprecht~134 & 92 & 86 & 45 & 74 & 51 & 43 & 59\\
Trumpler~5 & 563 & 562 & 78 & 349 & 141 & 75 & 114\\
Trumpler~20 & 287 & 149 & 115 & 209 & 185 & 179 & 164\\
Trumpler~23 & 30 & 20 & 13 & 17 & 15 & 13 & 15\\\hline
\end{tabular}
\end{table}

\section{Analysis and discussion}

To perform a comparative study across the cluster population, it was necessary to establish 
a consistent chemical baseline. We initially surveyed the selected 6050 stars sample distributed 
in 60 clusters to identify a subset of chemical species that were consistently measured across 
the largest number of open clusters. For the sake of homogeneity, we required that every open 
cluster in our study be analyzed using the same set of chemical species. This was done by 
simple removing chemical species without data, and then by keeping any chemical element with 
abundance measures in each of the 60 clusters, simultaneosly.

Furthermore, to ensure the statistical reliability of the open cluster fingerprint, we imposed 
a minimum occupancy constraint: each cluster-element pair must contain at least 10 stars with 
valid abundance measurements. This was done by simply keeping the chemical elements commonly 
found in the 60 clusters that have al least 10 stars per measured chemical element. This 
threshold was chosen to guarantee that the resulting mean abundance values and standard 
deviations are not dominated by small-number statistics or individual outliers.

This selection process resulted in a final sample of 34 open clusters and 7 chemical element 
abundances: [Fe/H], Li, Si, Ca, Ti, Co, and Ni (demonimation taken from GES catalogue). 
The total number of stars satisfying these 
criteria across the survey is 4914, providing a robust data set for investigating extra-tidal 
populations. To address the potential for chemical mimics --field stars that coincidentally 
share cluster-like abundances-- we consider the density of the field population in the 7-dimensional 
chemical space defined by [Fe/H], Li, Si, Ca, Ti, Co, and Ni. 
Given the high dimensionality of
this manifold, the chemical volume occupied by a single cluster is remarkably small, making 
accidental matches rare. While studies like \citet{casamiquelaetal2021} highlight the 
challenges of chemical overlap in the crowded Galactic disk even with 16 elements, our approach 
mitigates this by layering the chemical signature with spatial constraints. When this 
fingerprint is combined with the requirement that a star be located within or near the cluster's 
Jacobi radius, the probability of a field interloper matching both criteria becomes statistically 
negligible. Thus, while chemistry alone may not be a sufficient condition for membership 
in a blind survey, it acts as a powerful discriminant for stars already associated with the 
cluster's broad kinematic and spatial structure. The inclusion of Lithium (Li) is particularly 
significant here, as it provides 
an age-dependent chemical constraint \citep{,gutierrezetal2020,magrinietal2021} that helps break 
degeneracies between clusters and field stars that might 
otherwise share similar iron-peak or $\alpha$-element abundances. By requiring a simultaneous 
match across all seven dimensions, we drastically reduce the probability of a field star matching 
a cluster profile by pure chance.
Table~\ref{tab1} lists the clusters that have measures for the same chemical
elements, simultaneously, and the number of stars measured for each chemical species. As can 
be seen, every chemical element has more than 10 measured stars.

The core of our analysis strategy consists of obtaining a unique chemical fingerprint 
for each system, from which performing the chemical tagging step across the entire 
observed open cluster field. Since each star has an associated membership probability,
it is possible to spatially map the membership probabilities of stars with the
same chemical fingerprint. Both, chemical fingerprints and membership
probabilities were derived independently one to each other. The former comes from the
GES chemical element abundance values, while the latter comes from GES radial velocities
in combination with {\it Gaia} data.

The GES membership probabilities (\texttt{Mem3D}, hereafter $P$) used in this work are 
derived using a maximum-likelihood approach that combines spectral radial velocities with 
{\it Gaia} astrometry. As detailed in \citet{jacksonetal2022}, the GES pipeline models the 
distribution of stars in the cluster field as a combination of two populations: a cluster 
signal and a field background. Specifically, the membership is calculated in a 3D parameter 
space consisting of proper motions and radial velocities. The cluster members are assumed to 
follow a Gaussian distribution in these parameters, while the field is modeled using a broader, 
more complex distribution in proper motion and radial velocity. The resulting $P$ value 
represents the probability that a star's kinematics are consistent with the cluster core. 

In this study, we utilize stars with $P > 0.95$ from the 4914 stars in 34 clusters as our 
primary reference members to define the chemical signature, as this threshold 
minimizes field contamination to negligible levels. For each of the 34 selected open clusters
(see Table~\ref{tab1}), we calculated the mean abundance and dispersion for the 7 selected 
chemical abundance values using only the high-probability reference members ($P > 0.95$). We 
applied a sigma-clipping algorithm that removed any value beyond 3 sigma the initial mean value.
The mean and dispersion were recalculated until the values converged.
This procedure yielded a final table (see Table~\ref{tab2}) of reference chemical signatures 
(means and standard deviations) that represent the true chemical state of the cluster birth 
environment.

\begin{table*}
\caption{Mean values and dispersion of selected chemical species in open clusters.}
\label{tab2}
\begin{tabular}{lcccccccc}
\hline
Name & N & [Fe/H] & Li & Si & Ca &  Ti & Co & Ni \\
     & (dex) & (dex) & (dex) & (dex) & (dex) & (dex) & (dex) \\
\hline
Blanco~1 & 12&0.01 $\pm$ 0.06 & 1.49 $\pm$ 1.31 & 7.41 $\pm$ 0.05 & 6.39 $\pm$ 0.16 & 4.88 $\pm$ 0.12 & 4.81 $\pm$ 0.20 & 6.18 $\pm$ 0.12 \\
Berkeley~21 & 21 & -0.21 $\pm$ 0.21 & 1.71 $\pm$ 1.11 & 7.32 $\pm$ 0.13 & 6.22 $\pm$ 0.24 & 4.68 $\pm$ 0.17 & 4.70 $\pm$ 0.12 & 6.07 $\pm$ 0.18 \\
Berkeley~31 & 47 & -0.38 $\pm$ 0.14 & 1.73 $\pm$ 0.83 & 7.32 $\pm$ 0.26 & 6.19 $\pm$ 0.27 & 5.04 $\pm$ 0.35 & 5.13 $\pm$ 0.49 & 6.04 $\pm$ 0.29 \\
Berkeley~32 & 21 & -0.39 $\pm$ 0.11 & 2.01 $\pm$ 0.83 & 7.17 $\pm$ 0.16 & 6.12 $\pm$ 0.20 & 4.62 $\pm$ 0.36 & 4.66 $\pm$ 0.12 & 5.95 $\pm$ 0.15 \\
Berkeley~36 & 91 & -0.21 $\pm$ 0.13 & 1.64 $\pm$ 0.96 & 7.45 $\pm$ 0.26 & 6.29 $\pm$ 0.22 & 5.10 $\pm$ 0.42 & 5.22 $\pm$ 0.42 & 6.35 $\pm$ 0.28 \\
Berkeley~39 & 8 & -0.24 $\pm$ 0.12 & 2.30 $\pm$ 0.60 & 7.38 $\pm$ 0.04 & 6.19 $\pm$ 0.19 & 4.83 $\pm$ 0.13 & 4.81 $\pm$ 0.06 & 6.17 $\pm$ 0.12 \\
Berkeley~44 & 11& 0.12 $\pm$ 0.10 & 2.12 $\pm$ 1.09 & 7.53 $\pm$ 0.12 & 6.41 $\pm$ 0.20 & 5.02 $\pm$ 0.17 & 5.17 $\pm$ 0.27 & 6.50 $\pm$ 0.22 \\
Berkeley~81 &14 & 0.16 $\pm$ 0.17 & 1.65 $\pm$ 1.19 & 7.69 $\pm$ 0.12 & 6.36 $\pm$ 0.16 & 5.01 $\pm$ 0.25 & 5.32 $\pm$ 0.45 & 6.39 $\pm$ 0.30 \\
Haffner~10 & 15& -0.23 $\pm$ 0.15 & 2.36 $\pm$ 0.73 & 7.38 $\pm$ 0.04 & 6.32 $\pm$ 0.22 & 4.73 $\pm$ 0.11 & 4.74 $\pm$ 0.05 & 6.10 $\pm$ 0.16 \\
IC~2602 & 8& -0.04 $\pm$ 0.09 & 2.62 $\pm$ 1.09 & 7.39 $\pm$ 0.10 & 6.37 $\pm$ 0.16 & 4.85 $\pm$ 0.12 & 4.86 $\pm$ 0.27 & 6.11 $\pm$ 0.13 \\
NGC~2141 & 29& -0.11 $\pm$ 0.09 & 2.48 $\pm$ 0.70 & 7.43 $\pm$ 0.08 & 6.47 $\pm$ 0.24 & 4.75 $\pm$ 0.41 & 4.84 $\pm$ 0.10 & 6.18 $\pm$ 0.13 \\
NGC~2158 & 55 & -0.24 $\pm$ 0.18 & 2.24 $\pm$ 1.09 & 7.35 $\pm$ 0.12 & 6.29 $\pm$ 0.25 & 4.78 $\pm$ 0.41 & 4.75 $\pm$ 0.10 & 6.13 $\pm$ 0.12 \\
NGC~2243 & 287 &  -0.62 $\pm$ 0.14 & 2.40 $\pm$ 0.47 & 7.22 $\pm$ 0.27 & 6.00 $\pm$ 0.24 & 5.09 $\pm$ 0.35 & 5.16 $\pm$ 0.44 & 5.77 $\pm$ 0.25 \\
NGC~2264 & 10& -0.15 $\pm$ 0.16 & 3.56 $\pm$ 0.25 & 7.56 $\pm$ 0.19 & 6.32 $\pm$ 0.20 & 4.98 $\pm$ 0.26 & 4.93 $\pm$ 0.19 & 6.24 $\pm$ 0.22 \\
NGC~2420 & 228&  -0.24 $\pm$ 0.12 & 2.50 $\pm$ 0.53 & 7.35 $\pm$ 0.20 & 6.23 $\pm$ 0.18 & 5.09 $\pm$ 0.33 & 5.09 $\pm$ 0.29 & 6.05 $\pm$ 0.23 \\
NGC~2425 & 10&  -0.23 $\pm$ 0.12 & 2.47 $\pm$ 0.74 & 7.35 $\pm$ 0.09 & 6.29 $\pm$ 0.23 & 4.79 $\pm$ 0.22 & 4.76 $\pm$ 0.13 & 6.07 $\pm$ 0.25 \\
NGC~2516 & 115&-0.01 $\pm$ 0.07 & 1.43 $\pm$ 1.16 & 7.52 $\pm$ 0.24 & 6.18 $\pm$ 0.20 & 4.82 $\pm$ 0.12 & 4.90 $\pm$ 0.22 & 6.22 $\pm$ 0.14 \\
NGC~2547 &12 &-0.03 $\pm$ 0.13 & 2.49 $\pm$ 1.20 & 7.47 $\pm$ 0.14 & 6.34 $\pm$ 0.17 & 4.95 $\pm$ 0.22 & 4.88 $\pm$ 0.29 & 6.15 $\pm$ 0.17 \\
NGC~2682 & 81&-0.01 $\pm$ 0.06 & 1.59 $\pm$ 0.93 & 7.42 $\pm$ 0.06 & 6.26 $\pm$ 0.07 & 4.78 $\pm$ 0.18 & 4.90 $\pm$ 0.07 & 6.24 $\pm$ 0.07 \\
NGC~3532 & 46&-0.02 $\pm$ 0.12 & 1.65 $\pm$ 1.21 & 7.46 $\pm$ 0.13 & 6.28 $\pm$ 0.14 & 4.93 $\pm$ 0.21 & 4.86 $\pm$ 0.12 & 6.25 $\pm$ 0.15 \\
NGC~4815 & 12&0.08 $\pm$ 0.16 & 2.40 $\pm$ 1.05 & 7.56 $\pm$ 0.16 & 6.58 $\pm$ 0.78 & 5.21 $\pm$ 0.59 & 5.24 $\pm$ 0.49 & 6.27 $\pm$ 0.32 \\
NGC~6005 & 20&0.12 $\pm$ 0.14 & 1.79 $\pm$ 1.06 & 7.68 $\pm$ 0.10 & 6.51 $\pm$ 0.21 & 5.02 $\pm$ 0.23 & 5.22 $\pm$ 0.25 & 6.49 $\pm$ 0.21 \\
NGC~6067 & 14&0.07 $\pm$ 0.14 & 2.58 $\pm$ 1.20 & 7.58 $\pm$ 0.08 & 6.56 $\pm$ 0.38 & 5.35 $\pm$ 0.79 & 4.99 $\pm$ 0.34 & 6.48 $\pm$ 0.36 \\
NGC~6253 & 128&0.30 $\pm$ 0.10 & 1.65 $\pm$ 0.75 & 7.78 $\pm$ 0.10 & 6.52 $\pm$ 0.12 & 5.21 $\pm$ 0.18 & 5.33 $\pm$ 0.16 & 6.63 $\pm$ 0.12 \\
NGC~6259 & 16&0.16 $\pm$ 0.09 & 2.07 $\pm$ 1.00 & 7.75 $\pm$ 0.18 & 6.56 $\pm$ 0.24 & 5.08 $\pm$ 0.09 & 5.21 $\pm$ 0.17 & 6.46 $\pm$ 0.14 \\
NGC~6281 & 13&0.04 $\pm$ 0.13 & 0.85 $\pm$ 0.92 & 7.52 $\pm$ 0.14 & 6.29 $\pm$ 0.17 & 4.99 $\pm$ 0.19 & 4.90 $\pm$ 0.14 & 6.28 $\pm$ 0.09 \\
NGC~6405 &13 &-0.02 $\pm$ 0.12 & 2.10 $\pm$ 1.25 & 7.50 $\pm$ 0.14 & 6.45 $\pm$ 0.26 & 4.94 $\pm$ 0.39 & 4.78 $\pm$ 0.08 & 6.34 $\pm$ 0.16 \\
NGC~6633 & 7&-0.08 $\pm$ 0.13 & 0.80 $\pm$ 1.33 & 7.51 $\pm$ 0.22 & 6.26 $\pm$ 0.45 & 5.06 $\pm$ 0.62 & 4.95 $\pm$ 0.27 & 6.16 $\pm$ 0.18 \\
NGC~6705 &44 &0.06 $\pm$ 0.10 & 2.52 $\pm$ 0.75 & 7.61 $\pm$ 0.12 & 6.51 $\pm$ 0.26 & 4.95 $\pm$ 0.39 & 5.21 $\pm$ 0.40 & 6.37 $\pm$ 0.20 \\
NGC~6802 &15 &-0.01 $\pm$ 0.21 & 2.26 $\pm$ 0.86 & 7.59 $\pm$ 0.14 & 6.52 $\pm$ 0.22 & 5.01 $\pm$ 0.16 & 5.26 $\pm$ 0.54 & 6.27 $\pm$ 0.13 \\
Ruprecht~134 & 30&0.17 $\pm$ 0.10 & 2.24 $\pm$ 1.06 & 7.78 $\pm$ 0.13 & 6.74 $\pm$ 0.40 & 5.11 $\pm$ 0.32 & 5.35 $\pm$ 0.24 & 6.61 $\pm$ 0.22 \\
Trumpler~5 & 58&-0.47 $\pm$ 0.16 & 2.26 $\pm$ 0.73 & 7.15 $\pm$ 0.07 & 6.13 $\pm$ 0.23 & 4.56 $\pm$ 0.28 & 4.58 $\pm$ 0.06 & 5.93 $\pm$ 0.11 \\
Trumpler~20 &83 &0.06 $\pm$ 0.11 & 1.92 $\pm$ 1.31 & 7.61 $\pm$ 0.14 & 6.46 $\pm$ 0.25 & 5.12 $\pm$ 0.32 & 5.23 $\pm$ 0.34 & 6.34 $\pm$ 0.27 \\
Trumpler~23 & 11&0.16 $\pm$ 0.09 & 1.73 $\pm$ 0.91 & 7.70 $\pm$ 0.10 & 6.49 $\pm$ 0.29 & 5.07 $\pm$ 0.28 & 5.19 $\pm$ 0.23 & 6.41 $\pm$ 0.10 \\
\hline
\end{tabular}
\end{table*}

Using the information of Table~\ref{tab2} we tagged each of the 4914 stars (Table~\ref{tab1})
as chemically consistent with a cluster if it satisfies considering their individual 
uncertainties a simultaneous 7-chemical abundance match. We defined the inclusion criterion as:\\

$|$[Fe/H]$_{\rm star}$ - $<$[Fe/H]$_{\rm cls}$$>$$|$ $\le$ 3$\times$$\sigma$([Fe/H]$_{\rm cls}$)\\

and \\

$|$$X$$_{\rm star}$ - $<$$X$$_{\rm cls}$$>$$|$ $\le$ 3$\times$$\sigma$($X$$_{\rm cls}$).\\

\noindent Here $X$ represent the chemical elements Li, Si, Ca, Ti, Co, Ni
and the  ``star'' and ``cls'' subscripts refer to the values of the searched stars
in the GES database and the clusters'  mean values and dispersions in 
Table~\ref{tab2}.

By requiring an absolute match across all seven dimensions simultaneously, we significantly 
reduce the probability of false positives. This stringent multi-element requirement allows us 
to identify stars with the chemical signature of the cluster even when they are located well 
beyond the Jacobi radius or possess divergent kinematics. 
We emphasize that the present approach differs from blind searches in two critical ways.
First, this study utilizes supervised chemical tagging. Rather than attempting to discover 
clusters from an unknown field,  we test stars already observed 
within the specific GES cluster fields, and use high-probability members ($P > 0.95$) to define a known 
parent fingerprint for 34 specific systems. We then test if individual leaky stars --already 
observed in the vicinity of these clusters-- match that specific signature.
Second, we employ a seven-element manifold ([Fe/H], Li, Si, Ca, Ti, Co, and Ni) to define the
fingerprints; Li being a robust age-dependent chemical constraint. In total, 1692 stars 
match the above chemistry requirement.

In the kinematic analysis of open clusters, stars with a membership probability ($P$) below 0.7 
are frequently categorized as dubious members or field interlopers. This threshold is widely 
adopted in the literature \citep[e.g.,][]{jacksonetal2022} to minimize contamination in cluster 
samples. However, purely kinematic or statistical membership assignments can overlook genuine 
cluster members. Low membership probabilities can arise from several physical factors: stars 
may be in the process of escaping the cluster, they may be located in the low-density tidal 
debris, or they may be unresolved binary systems. In the case of binaries, the orbital motion 
can introduce a velocity offset (in either radial velocity or proper motion) from the cluster’s 
mean motion, leading to a statistically lower membership assignment despite the star's physical
association.

From the initial sample of 1692 stars whose multi-element chemical profiles are identical to 
their respective cluster fingerprints, we identified a subset of 63 stars -- distributed across 
22 distinct clusters -- that possess membership probabilities $P < 0.7$ (see Table~\ref{tab1append}). 
Given their high chemical consistency but low kinematic membership, these stars represent prime 
candidates for leaky cluster members or tidal tail constituents. To further characterize their 
physical association with their hosts, we performed a spatial analysis using the tidal boundaries
 defined by the Jacobi radii ($r_J$) provided in the recent catalogue by \citet{hr2024}.

\begin{figure}
\includegraphics[width=\columnwidth]{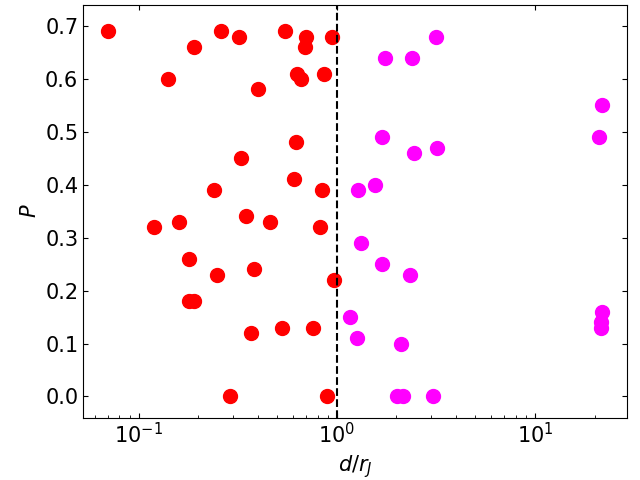}
\caption{Membership probabilities ($P$) as a function of $d$/$r_J$}
\label{fig1}
\end{figure}

The use of $r_J$ is critical for identifying stars that are no longer strictly 
bound by the cluster’s gravity but remain associated with its evolutionary history. Unlike 
empirical radii based on stellar density profiles (such as the King or Plummer radii), the 
Jacobi radius defines the boundary of the Roche lobe in the restricted three-body problem, 
specifically considering the cluster's mass and its position within the Milky Way potential.
In the work of \citet{hr2024}, these radii were derived using a comprehensive dynamical model 
applied to {\it Gaia} DR3 data. We used the stars' (RA, Dec.) coordinates and the central clusters'
positions and heliocentric distances \citep[][and references therein]{hr2024} to compute the star distances to the clusters' centres ($d$) and 
the ratio $d$/$r_J$. Table~\ref{tab1append} list the resulting values, while Figure~\ref{fig1}
illustrates the distribution of $P$ values in terms of  $d$/$r_J$. 

The multi-elemental chemical tagging identified a final sample of 63 stars across 22 open clusters 
that are chemically indistinguishable from the cluster mean but are kinematically dubious according 
to the {\it Gaia}-ESO membership pipeline ($P < 0.7$). It is important to note that the 63 stars 
identified here were not selected from a blind search of the entire {\it Gaia}-ESO catalogue. 
Rather, they represent a subset of stars that already passed initial kinematic filters (with 
$P > 0$). Therefore, the identification of these stars as recovered members relies on the joint 
probability of spatial, kinematic, and chemical coincidence. 
While \citet{casamiquelaetal2021} highlight the challenges of chemical overlap, recent results 
from the {\it Gaia}-ESO Survey \citep{spinaetal2022} indicate that specific data analysis 
strategies can successfully recover cluster members in abundance space.
By analyzing the spatial distribution of these 63 stars relative to their cluster's Jacobi 
radii, we distinguished between stars undergoing tidal stripping and those whose kinematic 
membership is masked by internal dynamics. Figure~\ref{fig1} shows that of the 63 stars, 
22 (35\%) are located in the extra-tidal region ($d > r_J$), while 41 (65\%) reside within 
the tidal boundary ($d \la r_J$). This bipartite distribution suggests that the failure of 
kinematic membership probabilities to capture these stars stems from two distinct physical 
phenomena: tidal evaporation into the Galactic field and internal velocity offsets.

The group of 22 stars is particularly noteworthy: despite their low kinematic membership probability, 
their chemical identity is an unambiguous fingerprint of their cluster origin. Their location 
outside the Jacobi radius strongly suggests they are part of the cluster's tidal tails or have 
recently leaked into the Milky Way field. This result confirms that stars in order to escape 
from the cluster need to reach velocities different from those of cluster's members. Then, 
the Milky Way potential imprints on them different accelerations, so that mean kinematic properties 
vary along tidal tail extensions \citep{piattietal2023,grondinetal2024}. This is an important
consideration since the identification of star cluster's tidal tail stars has been frequently 
addressed by looking for stars that are kinematically consistent with the mean kinematic 
properties of star clusters where the stars formed \citep{sollima2020,xuetal2024}. 
For the group of 41 stars, the low membership probability ($P < 0.7$) likely 
does not stem from escape, but rather from the kinematic noise. These are likely genuine members 
whose kinematic parameters have been shifted by binary orbital motion or that occupy the outer 
regions of the cluster where velocity dispersion are higher and more difficult to distinguish 
from the field population.

The most striking evidence of active cluster dissolution is found in Berkeley~31 and 
Ruprecht~134. Berkeley~31 hosts the largest subset of identified stars (15), of which 73\% 
(11 stars) are located outside the Jacobi radius. These stars exhibit a wide range of membership 
probabilities (0.00 $<$ $P$ $<$ 0.64), with an average $P \approx 0.25$. The significant 
presence of extra-tidal members suggests that Berkeley~31 is undergoing intense tidal stripping. 
The extremely low $P$ values for stars like 06581568+0823036 ($P$ = 0.00, $d$/$r_J$=3.05) indicate 
that kinematic membership models essentially give up on stars once they cross the tidal boundary, 
even if they remain chemically tied to the system.

Similarly, Ruprecht~134 shows a 100\% outside rate for its five identified stars, with distances 
exceeding 100 arcmin ($d$/$r_J \approx 21.5$). These are essentially lost members now populating 
the Milky Way field. The fact that their $P$ values range from 0.13 to 0.55 emphasizes that 
kinematic catalogues fail to recognize these stars as part of the cluster's extended tidal debris, 
whereas chemical tagging recovers them with high confidence.
NGC~6259 and NGC~6281 also show a 100\% $d$/$r_J > 1.0$ rate for their identified stars, 
suggesting they are in advanced stages of mass loss.

In contrast to the leaky systems, several clusters show a high concentration of chemically 
identical stars with 0.11 $<$ $P$ $<$ 0.69 located entirely inside the Jacobi radius. This group is 
led by NGC~6705, NGC~6253, and Trumpler~20, each containing 6 identified stars, all of which are 
spatially at $d$/$r_J < 1$.

The average membership probability in these clusters is notably higher than in the leaky clusters 
(e.g., $<$$P$$> \approx 0.54$ for NGC~6705). Because these stars are physically located in the 
cluster core or envelope, their low $P$ cannot be attributed to tidal escape. Instead, these are 
prime candidates for unresolved binary systems. In these cases, the orbital motion of the star 
around its companion introduces a velocity offset (either in radial velocity or proper motion) 
that deviates from the cluster's mean bulk motion. This kinematic noise causes the statistical 
membership pipelines \citep[e.g.,][]{jacksonetal2022} to penalize the star, despite it being a 
genuine physical member. Chemical tagging effectively bypasses this kinematic perturbation.

A unique case is found in NGC~6405, where a star (17400381-3214239) is located deep within 
the cluster ($d$/$r_J$=0.29) but has a membership probability of effectively zero ($P$ = 0.00). 
Such a high spatial-kinematic discrepancy is an evidence for a severe binary-induced velocity 
shift or a recent internal dynamical ejection event. Without the seven-element chemical fingerprint 
used in this study, such a star would be unequivocally rejected by any astrometric catalogue, leading 
to an underestimation of the cluster's mass and binary fraction.

The inclusion of stars with low kinematic membership probabilities is justified by their spatial 
proximity to the cluster centres and their position in the {\it Gaia}-ESO photometric survey. For these 
targets, the discrepancy in radial velocity is likely attributable to orbital motion in unresolved binary 
systems. However, we must consider the potential systematic effects of treating such systems with 
single-star spectroscopic models. As demonstrated by \citet{elbadryetal2018}, unresolved binaries can 
introduce biases in derived atmospheric parameters (typically $\sim$300 K in $T_{eff}$ and $\sim$0.1
dex in [Fe/H]) when fitted with single-star templates. We argue that these effects are minimized in our 
study for two reasons. First, the systematic biases are significantly less pronounced in the optical 
GIRAFFE/UVES spectra used here compared to near-infrared data. Second, the {\it Gaia} DR3 photometry 
provided in Table A.1 shows that these stars remain consistent with the cluster sequences, suggesting 
they are either single stars or have low mass ratios ($q < 0.6$) where the secondary's light contribution is 
negligible. The chemical consistency observed between these stars and high-probability members 
further suggests that any binary-induced systematics do not exceed our quoted observational uncertainties.

\section{Conclusions}

In this work, we have shown the efficacy of multi-elemental chemical tagging as a diagnostic 
tool for recovering the missing populations of open clusters. By utilizing high-resolution 
spectroscopy from the {\it Gaia}-ESO Survey (GES) and focusing on a sample of 63 stars across 
22 clusters that were previously dismissed by kinematic catalogues ($P$ $<$ 0.7), we have reached 
several key conclusions regarding the nature of cluster membership and dissolution in the 
Milky Way disk:\\

$\bullet$ The results confirm that standard membership 
probabilities based on {\it Gaia} astrometry are fundamentally conservative and spatially biased. 
While these catalogues are excellent for defining high-purity core samples, they systematically 
leak real members. The identification of 63 chemically identical stars with low $P$ values suggests 
that cluster membership is more extensive and complex than a single probability threshold can capture.\\

$\bullet$ We successfully identified 22 stars (35\% of the outlier sample) 
located outside the Jacobi radius. These stars, particularly those in open clusters like Berkeley~31 
and Ruprecht~134, provide direct empirical evidence of active cluster dissolution and the formation 
of tidal debris. The fact that these stars remain chemically indistinguishable from their host clusters
 while possessing near-zero kinematic membership probabilities highlights chemical tagging as the only 
reliable method for mapping the full extent of tidal tails in the Galactic field.\\

$\bullet$ For the 41 stars (65\% of the sample) located within the Jacobi radius, 
the low membership probabilities likely arise from kinematic noise induced by unresolved binary 
systems. Orbital motion creates velocity offsets that push these stars into the wings of the cluster’s 
phase-space distribution, causing them to be statistically penalized by maximum-likelihood algorithms. 
Our study shows that chemical tagging successfully bypasses these dynamical perturbations, allowing 
for a more accurate census of a cluster’s internal population and, potentially, its binary fraction.\\

$\bullet$ This study underscores the role of chemistry as a permanent birth certificate for stars. While a star's kinematics can be altered by tidal forces, dynamical encounters,
 or binary orbits, its elemental fingerprint ([Fe/H], Li, Si, Ca, Ti, Co, and Ni) remains preserved. It
 would seem that for future Galactic archaeology studies, the definition of cluster membership could evolve 
from a purely kinematic approach to a hybrid model that prioritizes chemical consistency.\\

In summary, the 63 stars recovered in this study represent the leaky components of the Milky Way’s 
open cluster population. By accounting for these stars, we move closer to a complete understanding 
of how stellar systems contribute to the assembly of the Milky Way disk. Future surveys combining 
high-resolution spectroscopy with the upcoming {\it Gaia} data releases will be essential to further bridge 
the gap between bound clusters and the field population.

\section*{Acknowledgements}
We thank the referee for the thorough reading of the manuscript and
timely suggestions to improve it.

\section{Data availability}
Data used in this work are available upon request to the first author.

%\newpage


%\bibliographystyle{mnras}
%\bibliography{paper} % if your bibtex file is called paper.bib

%to be uncommented before sending to editor
%\input{paper.bbl}

\newpage
\appendix
\onecolumn

\section{Selected open clusters' stars}

\begin{table}
\caption{$P$ $<$ 0.7 stars in the field of selected open clusters.}
\label{tab1append}
\begin{tabular}{lccccccccc}
\hline\hline
Name & star ID & RA & Dec. & $G$ & $G_{BP}$ & $G_{RP}$ & $P$ & $d$ & $d/r_J$ \\
 & & [deg] & [deg] & [mag] & [mag] & [mag] & & [arcmin] & \\
\hline
Berkeley~31 & 06581568+0823036 & 104.565300 & 8.384300 & 17.318 & 17.817 & 16.680 & 0.00 & 18.28 & 3.05 \\
Berkeley~31 & 06580232+0821538 & 104.509700 & 8.364900 & 17.176 & 17.671 & 16.530 & 0.46 & 14.78 & 2.46 \\
Berkeley~31 & 06580108+0820505 & 104.504500 & 8.347400 & 11.673 & 12.494 & 10.790 & 0.23 & 14.11 & 2.35 \\
Berkeley~31 & 06573833+0826375 & 104.409700 & 8.443800 & 16.849 & 17.232 & 16.318 & 0.00 & 12.96 & 2.16 \\
Berkeley~31 & 06575765+0817584 & 104.490200 & 8.299600 & 17.379 & 17.752 & 16.850 & 0.10 & 12.61 & 2.10 \\
Berkeley~31 & 06571827+0828000 & 104.326100 & 8.466700 & 16.917 & 17.296 & 16.382 & 0.00 & 12.11 & 2.02 \\
Berkeley~31 & 06570400+0826451 & 104.266700 & 8.445900 & 16.983 & 17.297 & 16.512 & 0.64 & 10.58 & 1.76 \\
Berkeley~31 & 06574658+0819057 & 104.444100 & 8.318300 & 17.396 & 17.761 & 16.912 & 0.25 & 10.16 & 1.69 \\
Berkeley~31 & 06573785+0818307 & 104.407700 & 8.308500 & 17.197 & 17.542 & 16.700 & 0.29 & 7.93 & 1.32 \\
Berkeley~31 & 06565536+0823122 & 104.230700 & 8.386700 & 15.554 & 16.098 & 14.865 & 0.11 & 7.59 & 1.27 \\
Berkeley~31 & 06573484+0817505 & 104.395200 & 8.297400 & 17.397 & 17.689 & 16.815 & 0.15 & 7.03 & 1.17 \\
Berkeley~31 & 06572487+0819135 & 104.353600 & 8.320400 & 16.887 & 17.341 & 16.276 & 0.00 & 5.32 & 0.89 \\
Berkeley~31 & 06571911+0820275 & 104.329600 & 8.341000 & 17.593 & 17.900 & 17.130 & 0.61 & 5.18 & 0.86 \\
Berkeley~31 & 06565453+0818165 & 104.227200 & 8.304600 & 16.688 & 17.031 & 16.187 & 0.61 & 3.76 & 0.63 \\
Berkeley~31 & 06570537+0813572 & 104.272400 & 8.232600 & 17.457 & 17.755 & 16.617 & 0.24 & 2.29 & 0.38 \\
Berkeley~36 & 07160485-1311211 & 109.020200 & -13.189200 & 17.716 & 18.653 & 16.783 & 0.39 & 6.41 & 1.28 \\
Berkeley~36 & 07162516-1311596 & 109.104800 & -13.199900 & 15.841 & 16.847 & 14.855 & 0.68 & 1.59 & 0.32 \\
Berkeley~44 & 19170646+1933119 & 289.276900 & 19.553300 & 18.074 & 19.248 & 17.031 & 0.26 & 0.98 & 0.18 \\
Berkeley~81 & 19015503-0030141 & 285.479300 & -0.503900 & 17.322 & 18.105 & 16.445 & 0.32 & 5.34 & 0.82 \\
IC~2602 & 10522362-6332566 & 163.098400 & -63.549100 & 10.696 & 11.101 & 10.118 & 0.47 & 80.17 & 3.21 \\
NGC~2158 & 06073722+2400536 & 91.905100 & 24.014900 & 14.421 & 15.238 & 13.532 & 0.13 & 5.67 & 0.76 \\
NGC~2243 & 06292602-3118247 & 97.358400 & -31.306900 & 16.339 & 16.601 & 15.924 & 0.33 & 1.96 & 0.16 \\
NGC~2264 & 06413250+0938074 & 100.385400 & 9.635400 & 14.396 & 15.093 & 13.512 & 0.68 & 17.01 & 0.95 \\
NGC~2420 & 07380419+2127598 & 114.517500 & 21.466600 & 17.686 & 17.924 & 16.715 & 0.22 & 8.24 & 0.97 \\
NGC~2420 & 07383259+2132068 & 114.635800 & 21.535200 & 15.869 & 16.096 & 15.365 & 0.45 & 2.85 & 0.33 \\
NGC~2516 & 07550546-6036336 & 118.772800 & -60.609300 & 13.894 & 14.392 & 13.227 & 0.49 & 23.66 & 1.69 \\
NGC~2516 & 07594001-6042311 & 119.916700 & -60.708600 & 14.414 & 15.003 & 13.671 & 0.39 & 11.76 & 0.84 \\
NGC~2516 & 07571050-6039362 & 119.293800 & -60.660100 & 13.505 & 13.972 & 12.870 & 0.48 & 8.67 & 0.62 \\
NGC~2682 & 08512080+1145024 & 132.836700 & 11.750700 & 12.947 & 13.253 & 12.479 & 0.66 & 3.90 & 0.19 \\
NGC~4815 & 12572603-6454596 & 194.358500 & -64.916600 & 17.933 & 18.752 & 17.103 & 0.33 & 2.51 & 0.46 \\
NGC~6005 & 15555468-5724358 & 238.977800 & -57.409900 & 16.173 & 16.752 & 15.345 & 0.69 & 2.84 & 0.55 \\
NGC~6005 & 15555171-5725390 & 238.965500 & -57.427500 & 13.431 & 14.209 & 12.539 & 0.12 & 1.90 & 0.37 \\
NGC~6005 & 15553294-5725298 & 238.887200 & -57.424900 & 13.856 & 14.701 & 12.945 & 0.23 & 1.32 & 0.25 \\
NGC~6067 & 16134290-5413206 & 243.428800 & -54.222400 & 16.337 & 16.872 & 15.572 & 0.58 & 3.82 & 0.40 \\
NGC~6067 & 16131806-5414582 & 243.325200 & -54.249500 & 16.522 & 17.009 & 15.759 & 0.18 & 1.78 & 0.19 \\
NGC~6253 & 16591359-5238458 & 254.806600 & -52.646100 & 16.510 & 17.039 & 15.819 & 0.41 & 4.54 & 0.61 \\
NGC~6253 & 16592701-5243289 & 254.862500 & -52.724700 & 14.481 & 15.114 & 13.712 & 0.13 & 3.95 & 0.53 \\
NGC~6253 & 16590933-5240330 & 254.789000 & -52.675900 & 15.277 & 15.744 & 14.636 & 0.34 & 2.65 & 0.35 \\
NGC~6253 & 16590787-5242569 & 254.782900 & -52.715900 & 15.452 & 15.922 & 14.812 & 0.60 & 1.02 & 0.14 \\
NGC~6253 & 16590681-5243165 & 254.778400 & -52.721200 & 15.344 & 15.810 & 14.698 & 0.32 & 0.93 & 0.12 \\
NGC~6253 & 16590371-5242318 & 254.765600 & -52.708900 & 15.511 & 15.961 & 14.869 & 0.69 & 0.53 & 0.07 \\
NGC~6259 & 17005877-4440503 & 255.244900 & -44.680600 & 15.986 & 16.606 & 15.075 & 0.68 & 25.45 & 3.18 \\
NGC~6259 & 17002391-4437579 & 255.099600 & -44.632800 & 16.577 & 17.264 & 15.761 & 0.64 & 19.22 & 2.40 \\
NGC~6281 & 17041286-3758022 & 256.053600 & -37.967300 & 14.758 & 15.313 & 14.051 & 0.40 & 9.66 & 1.56 \\
NGC~6405 & 17400381-3214239 & 265.015900 & -32.240000 & 13.995 & 14.506 & 13.303 & 0.00 & 3.47 & 0.29 \\
NGC~6705 & 18512384-0625366 & 282.849300 & -6.426800 & 15.719 & 16.303 & 14.970 & 0.68 & 10.48 & 0.70 \\
NGC~6705 & 18502885-0610475 & 282.620200 & -6.179900 & 16.482 & 17.028 & 15.773 & 0.66 & 10.34 & 0.69 \\
NGC~6705 & 18513103-0609074 & 282.879300 & -6.152100 & 16.131 & 16.704 & 15.401 & 0.60 & 9.92 & 0.66 \\
NGC~6705 & 18511731-0618154 & 282.822100 & -6.304300 & 16.381 & 16.937 & 15.520 & 0.69 & 3.84 & 0.26 \\
NGC~6705 & 18511808-0617223 & 282.825300 & -6.289500 & 16.654 & 17.220 & 15.939 & 0.39 & 3.67 & 0.24 \\
NGC~6705 & 18505637-0618193 & 282.734900 & -6.305400 & 16.993 & 17.414 & 16.009 & 0.18 & 2.69 & 0.18 \\
\hline
\end{tabular}
\end{table}

\setcounter{table}{0}
\begin{table}
\caption{continued.}
\label{tab1append}
\begin{tabular}{lccccccccc}
\hline\hline
Name & star ID & RA & Dec. & $G$ & $G_{BP}$ & $G_{RP}$ & $P$ & $d$ & $d/r_J$ \\
 & & [deg] & [deg] & [mag] & [mag] & [mag] & & [arcmin] & \\
\hline
Ruprecht~134 & 17523366-2929151 & 268.140200 & -29.487500 & 16.632 & --- & --- & 0.16 & 105.19 & 21.91 \\
Ruprecht~134 & 17525765-2931573 & 268.240200 & -29.532600 & 16.043 & 16.747 & 15.213 & 0.55 & 104.63 & 21.80 \\
Ruprecht~134 & 17522487-2930076 & 268.103600 & -29.502100 & 17.258 & 18.085 & 16.278 & 0.14 & 103.72 & 21.61 \\
Ruprecht~134 & 17530149-2933439 & 268.256200 & -29.562200 & 16.000 & 16.472 & 15.136 & 0.13 & 103.34 & 21.53 \\
Ruprecht~134 & 17525186-2935315 & 268.216100 & -29.592100 & 13.731 & 14.376 & 12.923 & 0.49 & 100.85 & 21.01 \\
Trumpler5 & 06364152+0928356 & 99.173000 & 9.476600 & 14.336 & 15.263 & 13.369 & 0.11 & 11.25 & 0.87 \\
Trumpler20 & 12385704-6040563 & 189.737700 & -60.682300 & 18.328 & 18.868 & 17.441 & 0.15 & 6.49 & 0.64 \\
Trumpler20 & 12400706-6037452 & 190.029400 & -60.629200 & 17.653 & 18.288 & 16.812 & 0.24 & 3.08 & 0.31 \\
Trumpler20 & 12392912-6035410 & 189.871300 & -60.594700 & 18.552 & 19.214 & 17.836 & 0.64 & 2.41 & 0.24 \\
Trumpler20 & 12392584-6038279 & 189.857700 & -60.641100 & 14.480 & 15.225 & 13.636 & 0.17 & 2.20 & 0.22 \\
Trumpler20 & 12395047-6036294 & 189.960300 & -60.608200 & 17.922 & 18.400 & 17.028 & 0.56 & 1.45 & 0.14 \\
Trumpler20 & 12393919-6038118 & 189.913300 & -60.636600 & 16.736 & 17.222 & 16.075 & 0.52 & 0.78 & 0.08 \\
\hline
\end{tabular}
\end{table}

\twocolumn

% Don't change these lines
\bsp	% typesetting comment
\label{lastpage}
\end{document}